\documentclass[conference]{IEEEtran}

\usepackage{xcolor}
\definecolor{links}{rgb}{0.5,0,0}   
\definecolor{urls}{rgb}{0,0,0.8}    
\definecolor{cites}{rgb}{0,0,0.6}   
\usepackage[colorlinks,hyperindex,linkcolor=links,citecolor=cites,urlcolor=urls]{hyperref} 

\usepackage{graphicx}
\usepackage{amsmath}
\usepackage{balance}
\usepackage{cite}
\usepackage{url}
\usepackage[inline]{enumitem} 

\usepackage{vmr-symbols-vecbold}
\usepackage{standard-macros}
\PassOptionsToPackage{hyphens}{url}
\usepackage{hyperref}

\DeclareSymbolFontAlphabet{\amsmathbb}{AMSb}%

\newcommand{\lro}[1]{\lefto({#1}\right)}																
\newcommand{\lrho}[1]{\lefto [ {#1} \right ]}																				

\newcommand{\lr}[1]{\left({#1}\right)}																
\newcommand{\lrb}[1]{\left \lbrace {#1} \right \rbrace}															

\safemath{\dopplerspread}{B_D}																								
\safemath{\delayspread}{T_D}																									
\safemath{\nc}{n\sub{c}}																										
\safemath{\nf}{n\sub{f}}																										
\safemath{\efa}{p\sub{sc}}
\safemath{\efb}{p\sub{cs}}
\safemath{\ef}{\epsilon\sub{f}	}
\safemath{\nd}{n\sub{d}}																										
\safemath{\ntx}{n\sub{t}} 																											
\safemath{\nrx}{n\sub{r}}																											
\safemath{\ntxt}{\tilde{n\sub{t}}}																											
\safemath{\cb}{\ensuremath{L}} 																								
\safemath{\cl}{\ensuremath{n}} 																								
\safemath{\txanto}{{\ensuremath{\tilde{m}_t}}} 																		
\safemath{\cs}{M} 																														
\safemath{\idPustm}{\ensuremath{S_{k}}}
\safemath{\error}{\ensuremath{\epsilon}} 																				
\safemath{\eexp}{\ensuremath{\mathcal{E}}} 																			
\safemath{\nsubc}{n\sub{s}}			 																						
\safemath{\nofdm}{n\sub{o}} 																									
\safemath{\bc}{\ensuremath{B_c}} 																							
\safemath{\ts}{\ensuremath{T_s}} 																							
\safemath{\nrb}{\ensuremath{n_{rb}}} 																						
\safemath{\rul}{\ensuremath{\rho\sub{ul}}}
\safemath{\rdl}{\ensuremath{\rho\sub{dl}}}

\safemath{\nres}{\ell}
\safemath{\nr}{n\sub{r}}
\safemath{\maxk}{M^*\lr{\nres, \nsubc, \nofdm, \epsilon, \rho}}
\safemath{\Rmax}{R^*}
\safemath{\Emin}{E\sub{b}^*/N_0}
\safemath{\Eminf}{\frac{E\sub{b}^*}{N_0}}
\safemath{\np}{\ensuremath{n\sub{p}}}
\safemath{\ndf}{\ensuremath{\bar{n}\sub{d}}}
\safemath{\npf}{\ensuremath{\bar{n}\sub{p}}}
\safemath{\code}{\ensuremath{\mathcal{C}}}
\safemath{\err}{\ensuremath{\epsilon}}
\safemath{\rp}{\ensuremath{\rho\sub{p}}}
\safemath{\rd}{\ensuremath{\rho\sub{d}}}
\safemath{\cohtime}{\ensuremath{T\sub{c}}}
\safemath{\cohbw}{\ensuremath{B\sub{c}}}
\safemath{\nmax}{\ensuremath{\ell\sub{m}}}
\safemath{\ntot}{\ensuremath{n\sub{tot}}}
\safemath{\nul}{\ensuremath{n\sub{ul}}}
\safemath{\ndl}{\ensuremath{n\sub{dl}}}

\safemath{\yp}{\ensuremath{\randvecy_{\nu}^{(\text{p})}}}
\safemath{\yd}{\ensuremath{\randvecy_{\nu}^{(\text{d})}}}
\safemath{\ypd}{\ensuremath{\vecy_{\nu}^{(\text{p})}}}
\safemath{\ydd}{\ensuremath{\vecy_{\nu}^{(\text{d})}}}

\safemath{\ypf}{\ensuremath{\bar{\randvecy}_{\nu}^{(\text{p})}}}
\safemath{\ydf}{\ensuremath{\bar{\randvecy}_{\nu}^{(\text{d})}}}
\safemath{\ypdf}{\ensuremath{\bar{\vecy}_{\nu}^{(\text{p})}}}
\safemath{\yddf}{\ensuremath{\bar{\vecy}_{\nu}^{(\text{d})}}}

\safemath{\xp}{\ensuremath{\vecx^{(\text{p})}}}
\safemath{\xd}{\ensuremath{\randvecx_{\nu}^{(\text{d})}}}
\safemath{\xdd}{\ensuremath{\vecx_{\nu}^{(\text{d})}}}

\safemath{\xpf}{\ensuremath{\bar{\vecx}^{(\text{p})}}}
\safemath{\xdf}{\ensuremath{\bar{\randvecx}_{\nu}^{(\text{d})}}}
\safemath{\xddf}{\ensuremath{\bar{\vecx}_{\nu}^{(\text{d})}}}

\safemath{\xdb}{\ensuremath{\overline{\randvecx}^{(\text{d})}}}
\safemath{\Pxd}{\ensuremath{P_{\randvecx^{(\text{d})}}}}

\safemath{\xpbar}{\ensuremath{\overline{\matX}^{(\text{p})}}}
\safemath{\xdbar}{\ensuremath{\overline{\randmatX}^{(\text{d})}}}

\safemath{\xdv}{\ensuremath{\randvecx^{(\text{d})}}}
\safemath{\xdbarv}{\ensuremath{\overline{\randvecx}^{(\text{d})}}}
\safemath{\ydv}{\ensuremath{\randvecy^{(\text{d})}}}

\safemath{\xdr}{\ensuremath{\matX^{(\text{d})}}}

\safemath{\ttx}{\ensuremath{\tau\sub{tx}}}
\safemath{\trx}{\ensuremath{\tau\sub{rx}}}
\safemath{\ack}{\ensuremath{\mathrm{s}}}
\safemath{\nack}{\ensuremath{\mathrm{c}}}

\newcommand{\prob}[1]{\ensuremath{\mathbb{P}\lrho{#1}}}

\safemath{\mI}{\ensuremath{i\lro{\randvecy ; \randvecx}}} 				

\safemath{\randveca}{\bm{A}}
\safemath{\randvecb}{\bm{B}}
\safemath{\randvecc}{\bm{C}}
\safemath{\randvecd}{\bm{D}}
\safemath{\randvece}{\bm{E}}
\safemath{\randvecf}{\bm{F}}
\safemath{\randvecg}{\bm{G}}
\safemath{\randvech}{\bm{H}}
\safemath{\randveci}{\bm{I}}
\safemath{\randvecj}{\bm{J}}
\safemath{\randveck}{\bm{K}}
\safemath{\randvecl}{\bm{L}}
\safemath{\randvecm}{\bm{M}}
\safemath{\randvecn}{\bm{N}}
\safemath{\randveco}{\bm{O}}
\safemath{\randvecp}{\bm{P}}
\safemath{\randvecq}{\bm{Q}}
\safemath{\randvecr}{\bm{R}}
\safemath{\randvecs}{\bm{S}}
\safemath{\randvect}{\bm{T}}
\safemath{\randvecu}{\bm{U}}
\safemath{\randvecv}{\bm{V}}
\safemath{\randvecw}{\bm{W}}
\safemath{\randvecx}{\bm{X}}
\safemath{\randvecy}{\bm{Y}}
\safemath{\randvecz}{\bm{Z}}
\safemath{\randvecphi}{\bm{\Phi}}

\safemath{\randmatA}{\amsmathbb{A}}
\safemath{\randmatB}{\amsmathbb{B}}
\safemath{\randmatC}{\amsmathbb{C}}
\safemath{\randmatD}{\amsmathbb{D}}
\safemath{\randmatE}{\amsmathbb{E}}
\safemath{\randmatF}{\amsmathbb{F}}
\safemath{\randmatG}{\amsmathbb{G}}
\safemath{\randmatH}{\amsmathbb{H}}
\safemath{\randmatI}{\amsmathbb{I}}
\safemath{\randmatJ}{\amsmathbb{J}}
\safemath{\randmatK}{\amsmathbb{K}}
\safemath{\randmatL}{\amsmathbb{L}}
\safemath{\randmatM}{\amsmathbb{M}}
\safemath{\randmatN}{\amsmathbb{N}}
\safemath{\randmatO}{\amsmathbb{O}}
\safemath{\randmatP}{\amsmathbb{P}}
\safemath{\randmatQ}{\amsmathbb{Q}}
\safemath{\randmatR}{\amsmathbb{R}}
\safemath{\randmatS}{\amsmathbb{S}}
\safemath{\randmatT}{\amsmathbb{T}}
\safemath{\randmatU}{\amsmathbb{U}}
\safemath{\randmatV}{\amsmathbb{V}}
\safemath{\randmatW}{\amsmathbb{W}}
\safemath{\randmatX}{\amsmathbb{X}}
\safemath{\randmatY}{\amsmathbb{Y}}
\safemath{\randmatZ}{\amsmathbb{Z}}
\safemath{\randmatSigma}{\mathbb{\Sigma}}
\safemath{\randmatPhi}{\mathbb{\Phi}}
\safemath{\randmatLambda}{\mathbb{\Lambda}}

\safemath{\matSigma}{\bm{\Sigma}}
\safemath{\matPhi}{\bm{\Phi}}
\safemath{\matLambda}{\bm{\Lambda}}

\usepackage{glossaries}
\glsdisablehyper
\newglossaryentry{simo}
{
  name={SIMO},
  description={single-input multiple-output},
  first={\glsentrydesc{simo} (\glsentrytext{simo})}
}

\newglossaryentry{mmse}
{
  name={MMSE},
  description={minimum mean-square error},
  first={\glsentrydesc{mmse} (\glsentrytext{mmse})}
}

\newglossaryentry{lmmse}
{
  name={LMMSE},
  description={linear minimum mean-square error},
  first={\glsentrydesc{lmmse} (\glsentrytext{lmmse})}
}

\newglossaryentry{mf}
{
  name={MF},
  description={matched filtering},
  first={\glsentrydesc{mf} (\glsentrytext{mf})}
}

\newglossaryentry{qlmmse}
{
  name={QLMMSE},
  description={quasilinear MMSE},
  first={\glsentrydesc{qlmmse} (\glsentrytext{qlmmse})}
}

\newglossaryentry{na}
{
  name={NA},
  description={normal approximation},
  first={\glsentrydesc{na} (\glsentrytext{na})},
  plural={NAs},
  descriptionplural={normal approximations},
  firstplural={\glsentrydescplural{na} (\glsentryplural{na})}
}

\newglossaryentry{zf}
{
  name={ZF},
  description={zero-forcing},
  first={\glsentrydesc{zf} (\glsentrytext{zf})}
}

\newglossaryentry{rzf}
{
  name={RZF},
  description={regularized zero-forcing},
  first={\glsentrydesc{rzf} (\glsentrytext{rzf})}
}

\newglossaryentry{sir}
{
  name={SIR},
  description={signal-to-interference ratio},
  first={\glsentrydesc{sir} (\glsentrytext{sir})}
}

\newglossaryentry{snr}
{
  name={SNR},
  description={signal-to-noise ratio},
  first={\glsentrydesc{snr} (\glsentrytext{snr})}
}

\newglossaryentry{mse}
{
  name={MSE},
  description={mean-squared error},
  first={\glsentrydesc{mse} (\glsentrytext{mse})}
}
\newglossaryentry{mmmse}
{
  name={M-MMSE},
  description={multicell MMSE},
  first={\glsentrydesc{mmmse} (\glsentrytext{mmmse})}
}

\newglossaryentry{ls}
{
  name={LS},
  description={least-squares},
  first={\glsentrydesc{ls} (\glsentrytext{ls})}
}

\newglossaryentry{mr}
{
  name={MR},
  description={maximum ratio},
  first={\glsentrydesc{mr} (\glsentrytext{mr})}
}

\newglossaryentry{ue}
{
  name={UE},
  description={user equipment},
  first={\glsentrydesc{ue} (\glsentrytext{ue})},
  plural={UEs},
  descriptionplural={user equipments},
  firstplural={\glsentrydescplural{ue} (\glsentryplural{ue})}
}

\newglossaryentry{ap}
{
  name={AP},
  description={access point},
  first={\glsentrydesc{ap} (\glsentrytext{ap})},
  plural={APs},
  descriptionplural={access points},
  firstplural={\glsentrydescplural{ap} (\glsentryplural{ap})}
}

\newglossaryentry{cpu}
{
  name={CPU},
  description={central processing unit},
  first={\glsentrydesc{cpu} (\glsentrytext{cpu})},
  plural={CPUs},
  descriptionplural={central processing units},
  firstplural={\glsentrydescplural{cpu} (\glsentryplural{cpu})}
}

\newglossaryentry{bs}
{
  name={BS},
  description={base station},
  first={\glsentrydesc{bs} (\glsentrytext{bs})},
  plural={BSs},
  descriptionplural={base stations},
  firstplural={\glsentrydescplural{bs} (\glsentryplural{bs})}
}

\newglossaryentry{ostbc}
{
  name={OSTBC},
  description={orthogonal space-time block code},
  first={\glsentrydesc{ostbc} (\glsentrytext{ostbc})},
  plural={OSTBCs},
  descriptionplural={orthogonal space-time block codes},
  firstplural={\glsentrydescplural{ostbc} (\glsentryplural{ostbc})}
}

\newglossaryentry{biawgn}
{
  name={bi-AWGN},
  description={binary-input additive white Gaussian noise},
  first={\glsentrydesc{biawgn} (\glsentrytext{biawgn})}
}

\newglossaryentry{flnf}
{
  name={FLNF},
  description={fixed-length no-feedback},
  first={\glsentrydesc{flnf} (\glsentrytext{flnf})}
}

\newglossaryentry{crc}
{
  name={CRC},
  description={cyclic redundancy check},
  first={\glsentrydesc{crc} (\glsentrytext{crc})}
}

\newglossaryentry{bsc}
{
  name={BSC},
  description={binary symmetric channel},
  first={\glsentrydesc{bsc} (\glsentrytext{bsc})}
  plural={BSCs},
  descriptionplural={binary symmetric channels},
  firstplural={\glsentrydescplural{bsc} (\glsentryplural{bsc})}
}

\newglossaryentry{bec}
{
  name={BEC},
  description={binary-erasure channel},
  first={\glsentrydesc{bec} (\glsentrytext{bec})}
}
\newglossaryentry{awgn}
{
  name={AWGN},
  description={additive white Gaussian noise},
  first={\glsentrydesc{awgn} (\glsentrytext{awgn})}
}
\newglossaryentry{3gpp}
{
  name={3GPP},
  description={3rd generation partnership project},
  first={\glsentrydesc{3gpp} (\glsentrytext{3gpp})}
}

\newglossaryentry{dl}
{
  name={DL},
  description={downlink},
  first={\glsentrydesc{dl} (\glsentrytext{dl})}
}

\newglossaryentry{LED}
{
  name={LED},
  description={light emitting diode},
  first={\glsentrydesc{LED} (\glsentrytext{LED})},
  plural={LEDs},
  descriptionplural={light emitting diodes},
  firstplural={\glsentrydescplural{LED} (\glsentryplural{LED})}
}

\newglossaryentry{lte}
{
  name={LTE},
  description={long term evolution},
  first={\glsentrydesc{lte} (\glsentrytext{lte})}
}
\newglossaryentry{ofdm}
{
  name={OFDM},
  description={orthogonal frequency-division multiplexing},
  first={\glsentrydesc{ofdm} (\glsentrytext{ofdm})}
}

\newglossaryentry{ul}
{
  name={UL},
  description={uplink},
  first={\glsentrydesc{ul} (\glsentrytext{ul})}
}
\newglossaryentry{rb}
{
  name={RB},
  description={resource block},
  first={\glsentrydesc{rb} (\glsentrytext{rb})},
  plural={RBs},
  descriptionplural={resource blocks},
  firstplural={\glsentrydescplural{rb} (\glsentryplural{rb})}
}
\newglossaryentry{cu}
{
  name={CU},
  description={channel use},
  first={\glsentrydesc{cu} (\glsentrytext{cu})},
  plural={CUs},
  descriptionplural={channel uses},
  firstplural={\glsentrydescplural{cu} (\glsentryplural{cu})}
}
\newglossaryentry{tti}
{
  name={TTI},
  description={transmission time interval},
  first={\glsentrydesc{tti} (\glsentrytext{tti})},
  plural={TTI's},
  descriptionplural={transmission time intervals},
  firstplural={\glsentrydescplural{tti} (\glsentryplural{tti})}
}
\newglossaryentry{iid}
{
  name={i.i.d.},
  description={independent and identically distributed},
  first={\glsentrydesc{iid} (\glsentrytext{iid})}
}
\newglossaryentry{psd}
{
  name={PSD},
  description={power spectral density},
  first={\glsentrydesc{psd} (\glsentrytext{psd})}
}

\newglossaryentry{mimo}
{
  name={MIMO},
  description={multiple-input multiple-output},
  first={\glsentrydesc{mimo} (\glsentrytext{mimo})}
}

\newglossaryentry{bler}
{
  name={BLER},
  description={block error probability},
  first={\glsentrydesc{bler} (\glsentrytext{bler})}
}
\newglossaryentry{mtc}
{
  name={MTC},
  description={machine-type communication},
  first={\glsentrydesc{mtc} (\glsentrytext{mtc})}
}
\newacronym{csi}{CSI}{channel state information}
\newglossaryentry{dvbt}
{
  name={DVB-T},
  description={terrestrial digital video broadcasting},
  first={\glsentrydesc{dvbt} (\glsentrytext{dvbt})}
}
\newglossaryentry{pat}
{
  name={PAT},
  description={pilot-assisted transmission},
  first={\glsentrydesc{pat} (\glsentrytext{pat})}
}
\newglossaryentry{ml}
{
  name={ML},
  description={maximum likelihood},
  first={\glsentrydesc{ml} (\glsentrytext{ml})}
}
\newglossaryentry{dt}
{
  name={DT},
  description={dependence-testing},
  first={\glsentrydesc{dt} (\glsentrytext{dt})}
}
\newglossaryentry{ustm}
{
  name={USTM},
  description={unitary space-time modulation},
  first={\glsentrydesc{ustm} (\glsentrytext{ustm})}
}
\newglossaryentry{siso}
{
  name={SISO},
  description={single-input single-output},
  first={\glsentrydesc{siso} (\glsentrytext{siso})}
}
\newglossaryentry{snn}
{
  name={SNN},
  description={scaled nearest-neighbor},
  first={\glsentrydesc{snn} (\glsentrytext{snn})}
}

\newglossaryentry{snnd}
{
  name={SNND},
  description={scaled nearest-neighbor decoding},
  first={\glsentrydesc{snnd} (\glsentrytext{snnd})}
}
\newglossaryentry{rcus}
{
  name={RCUs},
  description={random-coding union bound with parameter $s$},
  first={\glsentrydesc{rcus} (\glsentrytext{rcus})}
}
\newglossaryentry{osd}
{
  name={OSD},
  description={ordered statistics decoding},
  first={\glsentrydesc{osd} (\glsentrytext{osd})}
}
\newglossaryentry{llr}
{
  name={LLR},
  description={log-likelihood ratio},
  first={\glsentrydesc{llr} (\glsentrytext{llr})}
   plural={LLR's},
  descriptionplural={log-likelihood ratios},
  firstplural={\glsentrydescplural{llr} (\glsentryplural{llr})}
}
\newglossaryentry{qpsk}
{
  name={QPSK},
  description={quaternary phase shift keying},
  first={\glsentrydesc{qpsk} (\glsentrytext{qpsk})}
}
\newglossaryentry{nlos}
{
  name={NLOS},
  description={non-line-of-sight},
  first={\glsentrydesc{nlos} (\glsentrytext{nlos})}
}
\newglossaryentry{los}
{
  name={LOS},
  description={line-of-sight},
  first={\glsentrydesc{los} (\glsentrytext{los})}
}
\newglossaryentry{mgf}
{
  name={MGF},
  description={moment-generating function},
  first={\glsentrydesc{mgf} (\glsentrytext{mgf})}
}
\newglossaryentry{cgf}
{
  name={CGF},
  description={cumulant-generating function},
  first={\glsentrydesc{cgf} (\glsentrytext{cgf})},
  plural={CGFs},
  descriptionplural={cumulant-generating functions},
  firstplural={\glsentrydescplural{cgf} (\glsentryplural{cgf})}
}
\newglossaryentry{pdf}
{
  name={PDF},
  description={probability density function},
  first={\glsentrydesc{pdf} (\glsentrytext{pdf})},
   plural={PDF's},
  descriptionplural={probability density functions},
  firstplural={\glsentrydescplural{pdf} (\glsentryplural{pdf})}
}

\newglossaryentry{ccdf}
{
  name={CCDF},
  description={complementary cumulative distribution function},
  first={\glsentrydesc{ccdf} (\glsentrytext{ccdf})}
}

\newglossaryentry{cdf}
{
  name={CDF},
  description={cumulative distribution function},
  first={\glsentrydesc{cdf} (\glsentrytext{cdf})}
}

\newglossaryentry{gmi}
{
  name={GMI},
  description={generalized mutual information},
  first={\glsentrydesc{gmi} (\glsentrytext{gmi})}
}

\newglossaryentry{arq}
{
  name={ARQ},
  description={automatic repeat request},
  first={\glsentrydesc{arq} (\glsentrytext{arq})}
}

\newglossaryentry{harq}
{
  name={HARQ},
  description={hybrid automatic repeat request},
  first={\glsentrydesc{harq} (\glsentrytext{harq})}
}

\newglossaryentry{fbl}
{
  name={FBL-NF},
  description={fixed blocklength with no feedback},
  first={\glsentrydesc{fbl} (\glsentrytext{fbl})}
}
\newglossaryentry{ssnf}
{
  name={SS-NF},
  description={single-shot no-feedback},
  first={\glsentrydesc{ssnf} (\glsentrytext{ssnf})}
}

\newacronym{urllc}{URLLC}{ultra-reliable low-latency communications}

\newglossaryentry{vlsf}
{
  name={VLSF},
  description={Variable-length stop-feedback},
  first={\glsentrydesc{vlsf} (\glsentrytext{vlsf})}
}

\newglossaryentry{vlnsf}
{
  name={VLNSF},
  description={variable-length noisy stop feedback},
  first={\glsentrydesc{vlnsf} (\glsentrytext{vlnsf})}
}

\newglossaryentry{dmt}
{
  name={DMT},
  description={diversity-multiplexing tradeoff},
  first={\glsentrydesc{dmt} (\glsentrytext{dmt})}
}

\newglossaryentry{dmdt}
{
  name={DMDT},
  description={diversity-multiplexing-delay tradeoff},
  first={\glsentrydesc{dmdt} (\glsentrytext{dmdt})}
}

\newglossaryentry{tddt}
{
  name={TDDT},
  description={throughput-diversity-delay tradeoff},
  first={\glsentrydesc{tddt} (\glsentrytext{tddt})}
}

\newglossaryentry{tdd}
{
  name={TDD},
  description={time-division duplex},
  first={\glsentrydesc{tdd} (\glsentrytext{tdd})}
}

\newglossaryentry{stbc}
{
  name={STBC},
  description={space-time block-codes},
  first={\glsentrydesc{stbc} (\glsentrytext{stbc})},
  plural={STBC's},
  descriptionplural={space-time block-codes},
  firstplural={\glsentrydescplural{stbc} (\glsentryplural{stbc})}
}

\newglossaryentry{harqir}
{
  name={HARQ-IR},
  description={hybrid automatic repeat request with incremental redundancy},
  first={\glsentrydesc{harqir} (\glsentrytext{harqir})}
}

\newglossaryentry{harqcc}
{
  name={HARQ-CC},
  description={hybrid automatic repeat request with chase combining},
  first={\glsentrydesc{harqcc} (\glsentrytext{harqcc})}
}

\newglossaryentry{wp1}
{
  name={w.p.1.},
  description={with probability one},
  first={\glsentrydesc{wp1} (\glsentrytext{wp1})}
}

\newglossaryentry{vlff}
{
  name={VLFF},
  description={variable-length full feedback},
  first={\glsentrydesc{vlff} (\glsentrytext{vlff})}
}

\newglossaryentry{dmc}
{
  name={DMC},
  description={discrete memoryless channel},
  first={\glsentrydesc{dmc} (\glsentrytext{dmc})}
  plural={DMCs},
  descriptionplural={discrete memoryless channels},
  firstplural={\glsentrydescplural{dmc} (\glsentryplural{dmc})}
}

\newglossaryentry{dnn}
{
  name={DNN},
  description={deep neural network},
  first={\glsentrydesc{dnn} (\glsentrytext{dnn})}
  plural={DNNs},
  descriptionplural={deep neural networks},
  firstplural={\glsentrydescplural{dnn} (\glsentryplural{dnn})}
}

\newglossaryentry{nn}
{
  name={NN},
  description={neural network},
  first={\glsentrydesc{nn} (\glsentrytext{nn})}
  plural={NNs},
  descriptionplural={neural networks},
  firstplural={\glsentrydescplural{nn} (\glsentryplural{nn})}
}

\newglossaryentry{cnn}
{
  name={CNN},
  description={convolutional neural network},
  first={\glsentrydesc{cnn} (\glsentrytext{cnn})}
  plural={CNNs},
  descriptionplural={convolutional neural networks},
  firstplural={\glsentrydescplural{cnn} (\glsentryplural{cnn})}
}

\newglossaryentry{ber}
{
  name={BER},
  description={bit error rate},
  first={\glsentrydesc{ber} (\glsentrytext{ber})}
}

\newglossaryentry{scss}
{
  name={SCSS},
  description={\emph{single-channel} source separation},
  first={\glsentrydesc{scss} (\glsentrytext{scss})}
}

\newglossaryentry{fft}
{
  name={FFT},
  description={fast Fourier transform},
  first={\glsentrydesc{fft} (\glsentrytext{fft})}
}

\newglossaryentry{soi}
{
  name={SOI},
  description={signal of interest},
  first={\glsentrydesc{soi} (\glsentrytext{soi})}
}

\newglossaryentry{map}
{
  name={MAP},
  description={maximum \emph{a posteriori}},
  first={\glsentrydesc{map} (\glsentrytext{map})}
}

\newglossaryentry{usc}
{
  name={TDC},
  description={``temporal-diversity" condition},
  first={\glsentrydesc{usc} (\glsentrytext{usc})}
}

\newglossaryentry{qam}
{
  name={QAM},
  description={quadrature amplitude modulation},
  first={\glsentrydesc{qam} (\glsentrytext{qam})}
}

\newglossaryentry{mit}
{
  name={MIT},
  description={Massachusetts Institute of Technology},
  first={\glsentrydesc{mit} (\glsentrytext{mit})}
}

\newglossaryentry{rf}
{
  name={RF},
  description={radio-frequency},
  first={\glsentrydesc{rf} (\glsentrytext{rf})}
}
\usepackage{pgfplots}
\pgfplotsset{compat=1.14}
\usepgfplotslibrary{groupplots}
\usetikzlibrary{pgfplots.groupplots} 
\usepackage{subcaption}
\newtheorem{theorem}{Theorem}
\newtheorem{lemma}{Lemma}
\newtheorem{corollary}{Corollary}
\newtheorem{definition}{Definition}

\newcommand{\lmmse}{\text{\tiny LMMSE}}
\newcommand{\mmse}{\text{\tiny MMSE}}
\newcommand{\map}{\text{\tiny MAP}}
\newcommand{\qlmmse}{\text{\tiny MAP-QLMMSE}}
\newcommand{\cnn}{\text{\tiny CNN}}
\newcommand{\cnnqlmmse}{\text{\tiny CNN-QLMMSE}}

\newcommand{\sir}{\text{\tiny SIR}}
\newcommand{\snr}{\text{\tiny SNR}}
\newcommand{\corr}{\text{\tiny corr}}

\makeatletter
\newcommand\footnoteref[1]{\protected@xdef\@thefnmark{\ref{#1}}\@footnotemark}
\makeatother
\IEEEoverridecommandlockouts

\begin{document}
\title{Data-Driven Blind Synchronization and Interference Rejection for Digital Communication Signals}
\author{\IEEEauthorblockN{Alejandro Lancho\IEEEauthorrefmark{1}, Amir Weiss\IEEEauthorrefmark{1}, Gary C.F. Lee, \\ Jennifer Tang, Yuheng Bu, Yury Polyanskiy, and Gregory W. Wornell}
\IEEEauthorblockA{
Massachusetts Institute of Technology, Cambridge, MA, USA\\
Emails: \{lancho, amirwei, glcf411, jstang, buyuheng, ypol, gww\}@mit.edu} 
\thanks{Research was sponsored by the United States Air Force Research Laboratory and the United States Air Force Artificial Intelligence Accelerator and was accomplished under Cooperative Agreement Number FA8750-19-2-1000. The views and conclusions contained in this document are those of the authors and should not be interpreted as representing the official policies, either expressed or implied, of the United States Air Force or the U.S. Government. The U.S. Government is authorized to reproduce and distribute reprints for Government purposes notwithstanding any copyright notation herein. Alejandro Lancho has received funding from the European Union’s Horizon 2020 research and innovation programme under the Marie Sklodowska-Curie grant agreement No 101024432. G. C.F. Lee is supported by the National Science Scholarship from the Agency for Science, Technology and Research (A*STAR). This work is also supported by the National Science Foundation under Grant No CCF-2131115.
\IEEEauthorrefmark{1}These authors contributed equally to this work.}}
 \maketitle
 \sloppy 
\begin{abstract}
  We study the potential of data-driven deep learning methods for separation of two communication signals from an observation of their mixture. In particular, we assume knowledge on the generation process of one of the signals, dubbed signal of interest (SOI), and no knowledge on the generation process of the second signal, referred to as interference. This form of the single-channel source separation problem is also referred to as interference rejection. 
  We show that capturing high-resolution temporal structures (nonstationarities), which enables accurate synchronization to both the SOI and the interference, leads to substantial performance gains. With this key insight, we propose a domain-informed neural network (NN) design that is able to improve upon both ``off-the-shelf" NNs and classical detection and interference rejection methods, as demonstrated in our simulations. Our findings highlight the key role communication-specific domain knowledge plays in the development of data-driven approaches that hold the promise of unprecedented gains.
\end{abstract}
\section{Introduction}\label{sec:intro}
The proliferation of wireless devices is leading to an increasingly crowded radio spectrum, and consequently, spectrum sharing will be unavoidable~\cite{Hirzallah17,Naik21}. Thus, different wireless communication systems will coexist in the same frequency bands, thereby generating unintentional interferences among them. In order to maintain high reliability, separation of the overlapping signals from the received mixture will become an essential building block in such communication systems.

In the image and audio domains, machine learning techniques have been successfully applied for \emph{source separation}, e.g., \cite{nugraha2016multichannel}. These methods usually exploit domain knowledge relating to the signals' structures. For example, color features and local dependencies are useful for separating natural images~\cite{gandelsman2019double}, whereas time-frequency spectrogram masking methods are typically adopted for separating audio signals~\cite{huang2015joint}.

For communication signals, if the sources are separable in time and/or frequency, one can separate them via appropriate masking and classical filtering methods (see, e.g., \cite{amin1997interference}). The key challenge in this domain is the separation of signals overlapping in \emph{both} time and frequency when the receiver is equipped with a single antenna, which inherently implies there is no spatial diversity to be exploited. This problem is also referred to as \gls{scss}. In this case, standard approaches exploiting spatial diversity for blind source separation, such as~\cite{comon2010handbook,weiss2019maximum}, are irrelevant.

Various methods are available in the literature to perform \gls{scss} of digital communication signals. A common approach is maximum likelihood sequence estimation of the target signal, for which algorithms such as particle filtering \cite{tu2007particle} and per-surviving processing algorithms \cite{tu2008single} can be used. However, such methods require prior knowledge of the signal models, which in practice may not be known or available.

Perhaps a more realistic approach is to assume that only a dataset of the underlying communication signals is available. This can be obtained, for example, through direct/background recordings, or using high fidelity simulators (e.g., \cite{o2016radio}), allowing for a \emph{data-driven} approach. In this setup, \glspl{dnn} arise as a natural choice. This problem has been recently promoted by the ``RF Challenge''~\cite{rfchallenge}. 

In this paper, we study the data-driven \gls{scss} problem where two communication signals overlap in time and frequency, and the receiver is equipped with one single antenna. We consider a \gls{soi} whose generation process is known, and an interference signal with cyclic statistical properties  that are unknown \emph{a priori}---as is the case in standard protocols.\footnote{We only assume the cyclic period is known. In practice, provided a dataset of the respective signal, this parameter can be consistently estimated \cite{napolitano2016cyclostationarity}.} This problem is also referred to as \emph{interference rejection}. As a performance measure, we consider the \gls{ber}.  

\paragraph*{Contributions} We show that temporal nonstationarities of the signals constitute strong regularities that translate to better separation conditions. In particular, when such temporal structures exist, the notion of (time-)synchronization becomes not only sensible, but advantageous for separation. Based on our theoretical results that bind synchronization with MMSE optimal separation, we propose a data-driven \gls{dnn} approach that is \gls{ber}-superior to the classical methods of demodulation with \gls{mf} and interference rejection with \gls{lmmse} estimation of the \gls{soi}. Our proposed \glspl{dnn} architectures, which can incorporate explicit synchronization, are inspired by specific domain knowledge, relevant to digital communication signals.

\paragraph*{Notation}
We use lowercase letters with standard font and sans-serif font, e.g., $x$ and $\rndx$, to denote deterministic and random scalars, respectively. Similarly, we use $\vecx$ and $\rvecx$ for deterministic and random vectors, respectively; and $\matX$ and $\rmatX$ for deterministic and random matrices, respectively.
The uniform distribution over a set $\setS$ is denoted as ${\rm Unif}(\setS)$, and for $K\in\naturals$, we denote $\setS_{K}\triangleq\{1,\ldots,K\}$. 
For brevity, we refer to the complex normal distribution as Gaussian. We denote $\matC_{zw}\triangleq\Exop\left[\rvecz\herm\rvecw\right]\in\complexset^{N_z \times N_w}$ as the covariance matrix of $\rvecz\in\complexset^{N_z\times 1}$ and $\rvecw\in\complexset^{N_w\times 1}$ (specializing to $\matC_{zz}$ for $\rvecz=\rvecw$). 
%
\section{Problem Formulation}\label{sec:problem}
We consider the single-channel, baseband signal model of a noisy mixture of two sources, given by
\begin{equation}
    \rndy[n] = \rnds[n-\rndk_s] + \rho_{\sir}^{-1/2} \rndb[n-\rndk_b] + \rho_{\snr}^{-1/2} \rndw[n], \,\, n\in\integers, \label{eq:proc_model}
\end{equation}
where $\rnds[n], \rndb[n] \in \complexset$ are assumed to be  cyclostationary processes with known fundamental cyclic periods $K_s, K_b\in\naturals$, respectively; $\rndw[n] \in \complexset$ denotes additive white Gaussian noise, statistically independent of $\rnds[n]$ and $\rndb[n]$; and $\rho_{\sir}, \rho_{\snr}\in \reals_+$. We refer to the signal $\rnds[n]$ as the \gls{soi}, and to $\rndb[n]$ as interference. The variables $\rndk_s, \rndk_b\in \integers$ denote unknown (discrete) time-shifts with respect to the start of the cyclic periods of $\rnds[n]$ and $\rndb[n]$, respectively, where the start of the cyclic periods are chosen arbitrarily to be at $n=0$ without loss of generality. Hence, we assume that $\rndk_s\sim{\rm Unif}\left(\setS_{K_s}\right)$ and $\rndk_b\sim{\rm Unif}\left(\setS_{K_b}\right)$.

Let $\rvecy\triangleq [\rndy[1] \cdots \rndy[N]]^{\rm T}$, $\rvecs(\rndk_s)\triangleq [\rnds[1-\rndk_s] \cdots \rnds[N-\rndk_s]]^{\rm T}$, $\rvecb(\rndk_b)\triangleq [\rndb[1-\rndk_b] \cdots \rndb[N-\rndk_b]]^{\rm T}$, and $\rvecw\triangleq [\rndw[1] \cdots \rndw[N]]^{\rm T}$. Then, we may compactly write \eqref{eq:proc_model} for $N$ samples as
\begin{equation}
    \rvecy = \rvecs(\rndk_s) + \rho_{\sir}^{-1/2} \rvecb(\rndk_b) + \rho_{\snr}^{-1/2} \rvecw\in \complexset^{N\times 1}. \label{eq:vector_model}
\end{equation}
We further assume that $\rvecs(\rndk_s)$ and $\rvecb(\rndk_b)$ are statistically independent, which is a reasonable assumption in scenarios of unintentional interference, for which each source is not actively jamming or adapting to the other signals present in the environment. For simplicity of the exposition, we assume that $\rvecs(\rndk_s)$ and $\rvecb(\rndk_b)$ are zero-mean, unit-average-power, i.e., their (possibly time-varying) variance averages to~$1$. In this case, the parameters $\rho_{\sir}, \rho_{\snr}$ represent the \gls{sir} and \gls{snr} at the receiver, respectively. 

The goal is to produce an estimate of $\rvecs(\rndk_s)$ from $\rvecy$, denoted by $\widehat{\rvecs}$, so that given some metric $\ell$, the cost $\Exop[\ell(\widehat{\rvecs}, \rvecs(\rndk_s))]$ is minimized. This problem is referred to as \gls{scss}. 

As mentioned in Section~\ref{sec:intro}, we assume we do not have precise knowledge of the underlying distributions of the \gls{soi} and the interference. However, we assume the availability of a dataset of the signals and their respective time-shifts $(\rvecs(\rndk_s),\rndk_s)$ and $(\rvecb(\rndk_b),\rndk_b)$, allowing for a data-driven approach. Examples of such datasets can be found in~\cite{rfchallenge,rfdeepai}.

 %
 \section{The Gain in Synchronization to Interference}\label{sec:sync}
Before we present our approach to the \gls{scss} problem formulated in Section \ref{sec:problem}, we provide an analysis of an asymptotically optimal estimator of $\rvecs(\rndk_s)$ for the metric $\ell(\vecx,\vecz)\triangleq\|\vecx-\vecz\|_2^2$, which will shed light on key aspects in optimal separation and the role of \emph{synchronization to interference}. 

In this section, we assume that $\rnds[n]$ and $\rndb[n]$ are Gaussian processes, which is a reasonable assumption to model some communication signals, e.g., \cite{banelli2000theoretical}. In this case, we define\footnote{Since $\rndw[n]$ is white (and therefore stationary), $\rndw[n-\rndk_b]$ is also white, hence without loss of generality we may indeed define \eqref{equivalentnoise} with the shift $\rndk_b$.}
\begin{equation}\label{equivalentnoise}
    \rndv[n-\rndk_b]\triangleq \rho_{\sir}^{-1/2} \rndb[n-\rndk_b] + \rho_{\snr}^{-1/2} \rndw[n], \quad n\in\integers,
\end{equation}
such that $\rvecv(\rndk_b)\triangleq\tp{\left[\rndv[1-\rndk_b] \cdots \rndv[N-\rndk_b]\right]}\in\complexset^{N\times 1}$ is the ``equivalent noise", which, given $\rndk_b$, is distributed as $\mathcal{CN}(\veczero,\matC_{vv})$. Thus, \eqref{eq:vector_model} simplifies to
\begin{equation}\label{compactmixture}
    \rvecy = \rvecs(\rndk_s) + \rvecv(\rndk_b)\in\complexset^{N\times 1}.
\end{equation}
Note that, generally, the equivalent noise term $\rvecv(\rndk_b)$ is not temporally white (as opposed to $\rvecw$), and exhibits a potentially informative statistical structure (e.g., in the form of $\matC_{vv}$) that can be exploited for enhanced separation performance.

\subsection{Linear \gls{mmse} Estimation}\label{sec:lmmseestimation}
A computationally attractive approach, which already exploits (some of) the underlying statistics of both of the components of the mixture \eqref{compactmixture}, is optimal \emph{linear} estimation. The \gls{lmmse} estimator \cite{van2004detection}, given by (assuming $\det(\matC_{yy})\neq0$)
\begin{equation}\label{lmmsedef}
    \widehat{\rvecs}_{\lmmse}\triangleq \matC_{sy}\inv{\matC_{yy}}\rvecy=\matC_{ss}\inv{\left(\matC_{ss}+\matC_{vv}\right)}\rvecy\in\complexset^{N\times 1},
\end{equation}
is constructed using the statistics of the mixture that inherently takes into account the potentially non-trivial structure of $\matC_{vv}$, i.e., some form of deviation from a scaled identity matrix.

However, while \eqref{lmmsedef} coincides with the \gls{mmse} estimator for jointly Gaussian processes, it is generally suboptimal due to the linearity constraint. Specifically, in our case, although the processes $\rnds[n], \rndv[n]$ are jointly Gaussian, $\rvecs(\rndk_s)$ and $\rvecv(\rndk_b)$ are \emph{not} even marginally Gaussian. Indeed, $\rvecs(\rndk_s)$ and $\rvecv(\rndk_b)$ are Gaussian mixtures due to the random time-shifts $\rndk_s,\rndk_b$. It then follows that \eqref{lmmsedef} is in fact not optimal, as shown next.

\subsection{\gls{mmse} Estimation}\label{subsec:mmseestimation}
The optimal estimator in the \gls{mmse} sense is known to be the conditional expectation,
\begin{equation}\label{mmsedefest}
    \widehat{\rvecs}_{\mmse}\triangleq\Exop[\rvecs(\rndk_s)|\rvecy]\in\complexset^{N\times 1},
\end{equation}
whose \gls{mse} is an achievable lower bound of the \gls{mse} of \emph{any} estimator of $\rvecs(\rndk_s)$. However, in most practical cases, \eqref{mmsedefest} is hard to obtain analytically and computationally. In our case, by using the law of total expectation in \eqref{mmsedefest}, the MMSE estimator is given by the more explicit and convenient form
\begin{IEEEeqnarray}{lCl}
\widehat{\rvecs}_{\mmse}&=&\Exop\left[\Exop[\rvecs(\rndk_s)|\rvecy,\rndk_s,\rndk_b] |\rvecy \right]\overset{(\star)}{=}\Exop\left[ \widehat{\rvecs}_{\lmmse}(\rndk_s,\rndk_b) |\rvecy \right]\nonumber\\
&=&\sum_{m_s=1}^{K_s}\sum_{m_b=1}^{K_b}\prob{\rndk_s=m_s,\rndk_b=m_b | \rvecy}\widehat{\rvecs}_{\lmmse}(m_s,m_b),\label{ourmmse}\IEEEeqnarraynumspace
\end{IEEEeqnarray}
where in $(\star)$ we have used the fact that, given the time-shifts, $\rvecs(\rndk_s)$ and $\rvecy$ are jointly Gaussian, and where $\widehat{\rvecs}_{\lmmse}(m_s,m_b)\triangleq\matC_{ss}(m_s)\inv{\left[\matC_{ss}(m_s)+\matC_{vv}(m_b)\right]}\rvecy$, with
\begin{equation}
\matC_{ss}(m)\triangleq\Exop[\rvecs\herm{\rvecs} | \rndk_s=m],\, \matC_{vv}(m)\triangleq\Exop[\rvecv\herm{\rvecv} | \rndk_b=m].\label{alignedstatistics}
\end{equation}
Put simply, \eqref{ourmmse} is a weighted average of $K_s \times K_b$ linear estimators, with the posterior probabilities---which are \emph{nonlinear} functions of the data $\rvecy$---serving as the normalized weights. Even before taking into account the computation of the posteriors, the sum in \eqref{ourmmse} scales with the product of possible time-shifts $K_s \times K_b$, rendering $\widehat{\rvecs}_{\mmse}$ often impractical.

As can be seen from \eqref{ourmmse}, synchronization (i.e., knowledge of the time-shifts) already substantially simplifies the computation, since, in that case,  only the (conditional) linear estimator $\widehat{\rvecs}_{\lmmse}(m_s,m_b)$ is required. In other words, eliminating this type of randomness from the mixture $\rvecy$ grants us lower computational complexity and a simple form of a linear estimator. Fortunately, a two-step ``synchronization-separation" estimator can approach the \gls{mmse} estimator, thus enjoying asymptotic optimality at a substantially reduced computational burden.

To show this rigorously, for simplicity of the exposition, we assume hereafter (unless stated otherwise) that the receiver is synchronized to the \gls{soi},\footnote{This is a reasonable assumption in most communication systems \cite{gao2015robust}.} namely, $\rndk_s=0$ and known. However, the result below can be generalized to the case where the \gls{soi}'s time-shift $\rndk_s$ is random and unknown. Let
\begin{equation}\label{mapestofkb}
     \widehat{\rndk}^{\map}_{b}\triangleq \argmax_{m\in\setS_{K_b}} \prob{\rndk_b=m|\rvecy}
\end{equation}
be the \gls{map} estimator of $\rndk_s$, and define the (suboptimal) ``plug-in", \gls{map}-based \emph{quasi}-linear \gls{mmse} estimator
\begin{equation}\label{mapqlmmse}
    \widehat{\rvecs}_{\qlmmse}\triangleq\widehat{\rvecs}_{\lmmse}(\widehat{\rndk}^{\map}_{b})\in\complexset^{N\times 1},
\end{equation}
where, for brevity, we use $\widehat{\rvecs}_{\lmmse}(m)$ to denote $\widehat{\rvecs}_{\lmmse}(0,m)$. Furthermore, we define the \gls{mse}s, as a function of $N$, as
\begin{align}
    \varepsilon^2_{\mmse}(N)&\triangleq\Exop\left[\|\widehat{\rvecs}_{\mmse}-\rvecs\|_2^2\right]\in\reals_+,\label{mmsedef}\\
    \varepsilon^2_{\qlmmse}(N)&\triangleq\Exop[\|\widehat{\rvecs}_{\qlmmse}-\rvecs\|_2^2]\in\reals_+\label{qlmmsedef}.
\end{align}

We now introduce a \gls{usc} under which optimal synchronization is increasingly accurate. 
\begin{definition}[\gls{usc}]\label{definition1}
Let $\psi_N(\rvecy,k)\triangleq\frac{1}{N}\herm{\rvecy}\matC^{-1}_{yy}(k)\rvecy -1$. The (sufficient) \gls{usc} is satisfied if there does not exist $k\in\setS_{K_b} \backslash \rndk_b$ such that $\lim_{N\to\infty} |\psi_N(\rvecy,k)|=0$.
\end{definition}
\begin{lemma}\label{lemma1}
Under the \gls{usc}, for any finite $\alpha\in\reals_+$, 
\begin{equation}\label{polydecayoferrprob}
    \prob{\widehat{\rndk}^{\map}_{b}\neq\rndk_{b}}=o\left(\frac{1}{N^{\alpha}}\right).
\end{equation}
\end{lemma}
\begin{IEEEproof}
See Appendix~\ref{app:proof_lemma1}.
\end{IEEEproof}
The theorem below shows that the two-step synchronization-separation approach \eqref{mapqlmmse} is asymptotically optimal.
\begin{theorem}\label{theorem1}
Under the \gls{usc}, we have
\begin{equation}\label{mseasymptoticequivalence}
\lim_{N\to\infty}\frac{\varepsilon^2_{\mmse}(N)}{\varepsilon^2_{\qlmmse}(N)}=1. 
\end{equation}
\end{theorem}
\begin{IEEEproof}
See Appendix \ref{app:proof_th}.
\end{IEEEproof}
In words, Theorem \ref{theorem1} tells us that, when the time-shift can be uniquely detectable, first optimally synchronizing to the interference, and then using a suboptimal, quasi-linear estimator, is asymptotically equivalent to \gls{mmse} estimation. Further intuition to this type of behaviour, for maximum-likelihood-based \gls{mmse} estimation, is given in \cite[Fig.~1]{weiss2019maximum}.

\subsection{Synchronization via \glspl{cnn}}\label{sec:synchviacnn}
\begin{figure}[t!]
\centering
\includegraphics[width=\columnwidth]{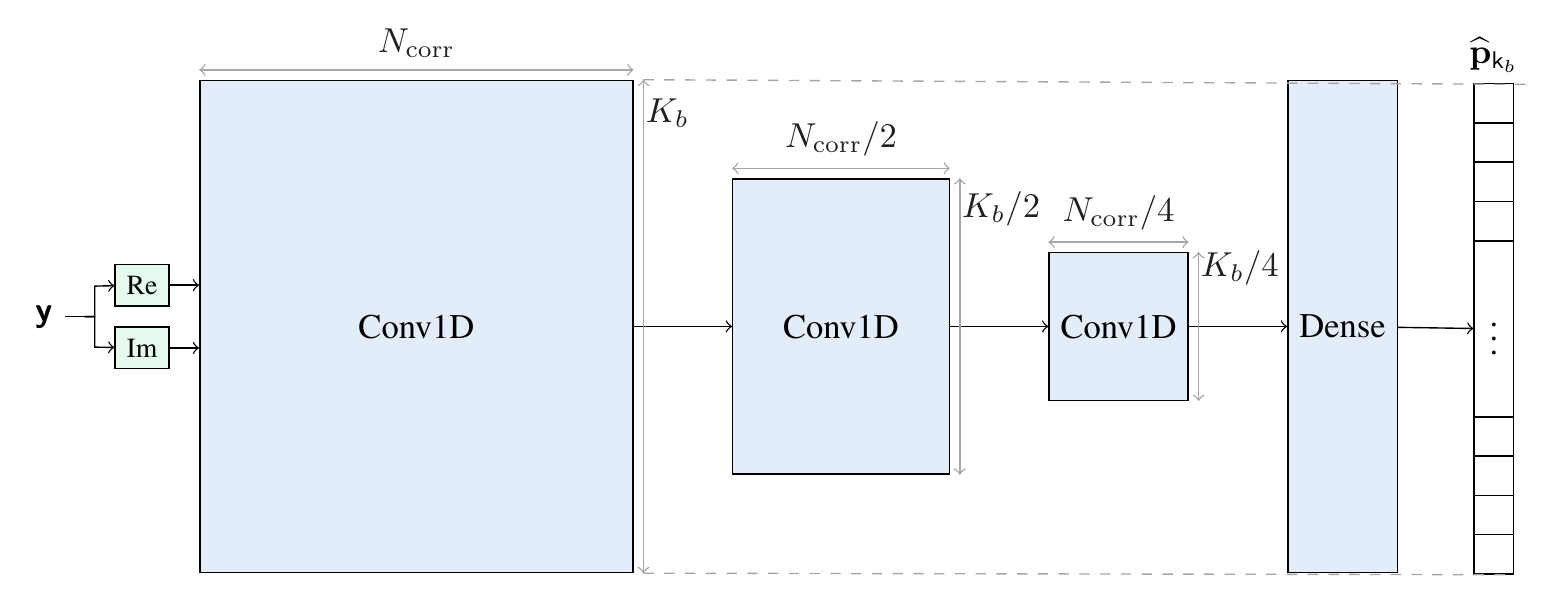}
\caption{Architecture of the proposed \gls{cnn} for synchronization.}
\label{fig:cnn_sync}
\end{figure}
Although the estimator \eqref{mapqlmmse} is attractive in terms of its \gls{mse} performance, it nevertheless requires---both for synchronization and separation---\emph{precise} knowledge of the underlying statistics, including the \gls{sir} and \gls{snr}, which can be hard to obtain in practice. In particular, without these statistics, it is impossible to obtain $\widehat{\rndk}_b^{\map}$. However, when by measurement or generation, sufficiently large datasets with realizations of $\rvecs(\rndk_s)$ and $\rvecb(\rndk_b)$ are available, a data-driven approach can be taken.

To this end, we leverage the strong capabilities of \gls{cnn}s for capturing intricate temporal structures, to train a synchronizer in a data-driven manner. Specifically, we propose the \gls{cnn}-based architecture depicted in Fig.~\ref{fig:cnn_sync}, which is trained in a supervised manner based on a labeled dataset of mixtures and the underlying interference time-shifts, $\{(\rvecy^{(i)}, \rndk_b^{(i)}): i \in \setS_{I_T}\}$, where $I_T$ is the size of the training dataset. We use a sufficiently large kernel size in the convolutional layers, which is proportional to the ``effective correlation length"---denoted as $N_\corr$ in Fig.~\ref{fig:cnn_sync}---so as to be able to capture the strongest, most informative temporal structures for estimation. Since the cyclic period $K_b$ is assumed to be known, we train a model using the cross-entropy loss, which receives as its input the mixture $\rvecy$ and outputs a vector of probabilities, denoted by $\widehat{\vecp}_{\rndk_b}\in[0,1]^{K_b\times 1}$. At inference time, we synchronize to the interference via $\widehat{\rndk}^{\cnn}_b\triangleq\argmax_{m\in\setS_{K_b}}\tp{\vece_m}\widehat{\vecp}_{\rndk_b}$ (cf.~$\widehat{\rndk}^{\map}_{b}$ in \eqref{mapestofkb}), where $\vece_m\in\reals^{K_b\times 1}$ denotes the $m$-th standard basis vector.

\begin{figure}[t!]     
    \centering
    \includegraphics[width=\columnwidth]{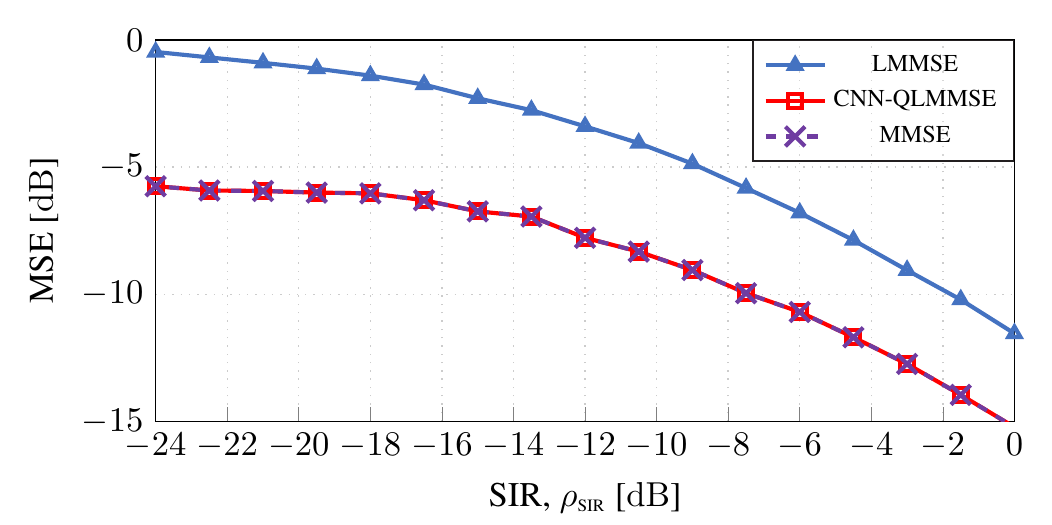}
    \caption{\gls{mse} as a function of the \gls{sir} ($\rho_{\sir}$) for a fixed \gls{snr} ($\rho_{\snr}$) of $20$\dB.}
    \label{fig:MSE_sync} 
\end{figure} 
In Fig.~\ref{fig:MSE_sync}, we show the \gls{mse} for $\widehat{\rvecs}_{\lmmse}, \widehat{\rvecs}_{\mmse}$ and 
\begin{equation}
\widehat{\rvecs}_{\cnnqlmmse}\triangleq\widehat{\rvecs}_{\lmmse}(\widehat{\rndk}^{\cnn}_{b}) \label{eq:cnn-qlmmse},  
\end{equation}
for the same communication waveforms described in detail in Section~\ref{sec:num}, but considering here Gaussian alphabets instead of the discrete and finite alphabets used in Section~\ref{sec:num}. 
As seen, the linearity restriction \eqref{lmmsedef} costs a considerable price in terms of the compromised performance relative to the lower bound, given by the \gls{mmse}. It is also evident that the \gls{mse} of the \gls{cnn}-based \gls{qlmmse} estimator $\widehat{\rvecs}_{\cnnqlmmse}$ coincides with \eqref{mmsedef}, which asymptotically coincides with the \gls{map}-\gls{qlmmse} \eqref{qlmmsedef} by virtue of Theorem~\ref{theorem1}.


All the above motivates our solution approach, and provides the theoretical foundations (as well as intuition) based on which we develop our system architecture, presented next.

 
 %
 \section{Interference Rejection via DNNs}\label{sec:int_rej_DNNs}
\begin{figure}[t!]
\centering
\includegraphics[width=\columnwidth]{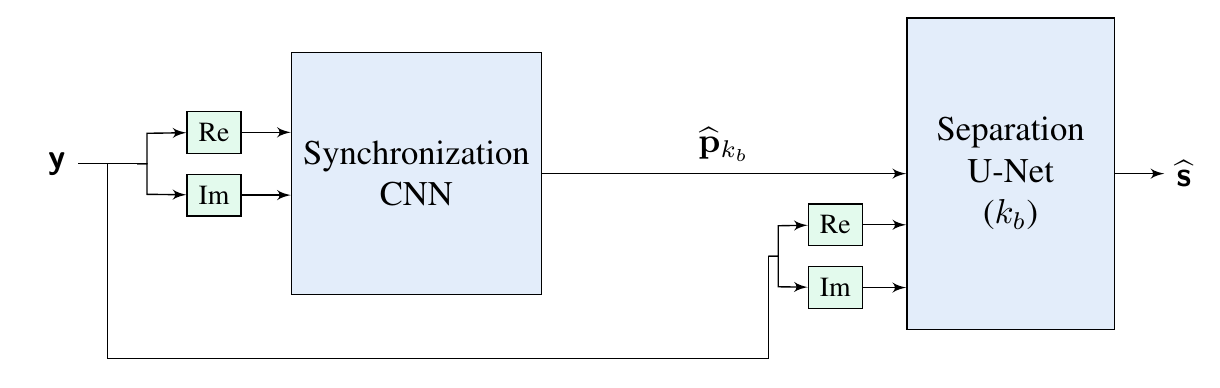}
\caption{System architecture of the \gls{dnn}--based approach with an explicit \gls{cnn}-based synchronization block prior to the separation block (U-Net).}
\label{fig:system}
\end{figure}

 We now present two supervised learning approaches for \gls{scss}, used in this work as interference rejection methods. The first \gls{dnn} architecture, depicted in Fig.~\ref{fig:system}, consists of two main building blocks: 
\begin{enumerate*}[label=(\roman*)]
  \item \gls{cnn} to perform synchronization to the interference,
  \item \gls{dnn} (U-Net) to perform \gls{scss}.\footnote{To separate the communication signals used in this paper, other \glspl{dnn} were implemented, yielding worse performance. Details can be found in the Github repository: \url{https://github.com/RFChallenge/SCSS_DNN_Comparison}.}
\end{enumerate*}
The key motivation to perform explicit synchronization is twofold. First, as explained in Section~\ref{subsec:mmseestimation}, due to Theorem~\ref{theorem1}, explicit consistent synchronization decoupled from separation, although suboptimal, can asymptotically (as $N\to\infty$) lead to optimal separation with reduced complexity. 
Second, although a sufficiently rich \gls{dnn} might be able to perform the synchronization and separation tasks jointly, for a given architecture, acquiring synchronization knowledge explicitly helps by reducing the complexity of the separation task. In Section~\ref{sec:expl_sync_num}, we show that this decoupled approach can indeed lead to performance gains. However, Lemma~\ref{lemma1} shows that there exists a realizable synchronization method that becomes increasingly accurate as the input size grows. While this can be exploited for explicit synchronization (e.g., Fig.~\ref{fig:system}), it could also imply that, under certain conditions, a \gls{dnn} architecture would be able to “implicitly synchronize” and separate, namely superior performance would be achieved \emph{without} explicit synchronization. This is shown in Section~\ref{sec:impl_high_dim}.
\begin{figure*}[t!]
\centering
\includegraphics[width=\textwidth]{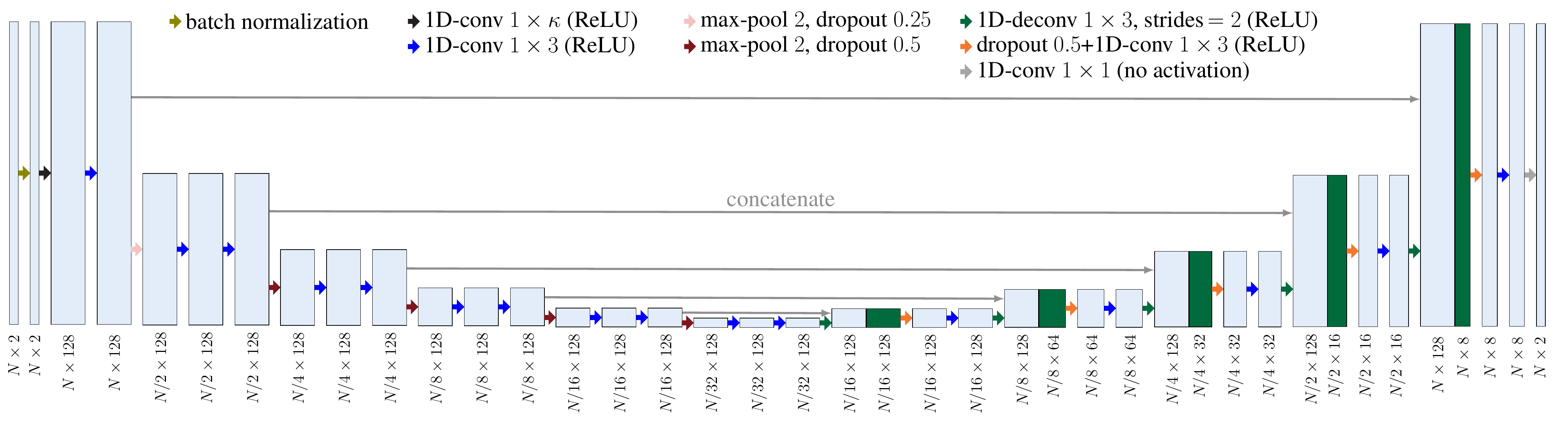}
\caption{Architecture of the \gls{dnn} (U-Net) proposed to perform \gls{scss} of the communication signals. The parameter $\kappa$ denotes the kernel size of the first layer.}
\label{fig:unet}
\end{figure*}

The synchronization block is based on the \gls{cnn} described in Section~\ref{sec:synchviacnn} (Fig.~\ref{fig:cnn_sync}). The \gls{dnn} for separation is based on the so-called U-Net (see Fig.~\ref{fig:unet}) 
\cite{ronneberger2015unet}, which has some properties that makes it suitable to the specific informative features of digital communication signals. In particular, its \gls{cnn} building blocks allow us to input and process long time intervals (e.g., $N> 10^4$), which cannot be processed using classical methods. In turn, processing such long signals allows for exploitation of temporal structures on a different scale, which can (and does) lead to substantial performance gains.

As shown in Fig.~\ref{fig:unet}, our \gls{dnn} approach departs from standard implementations intended to deal with images (2D signals). To handle 1D complex-valued, time-series  communication signals, we use 1D convolutional layers. Furthermore, differently from standard \gls{cnn}-based architectures that are designed to deal with images and hence use short kernels of size~$\sim\hspace{-0.05cm}3$ in all layers, our U-Net architecture utilizes a sufficiently long kernel in the first convolutional layer (denoted by $\kappa$ in Fig.~\ref{fig:unet}). This enables to capture the most influential temporal structures of the \gls{soi} and interference, which can lead to an order of magnitude gains, as demonstrated below.   

For training, we input the stacked real and imaginary parts of $\rvecy$ as separate channels to both the synchronization-to-interference \gls{cnn} and the separation U-Net. For separation, if explicit synchronization is performed, we mimic a nonlinear version of~\eqref{mapqlmmse} by using an instance of the \gls{dnn} architecture depicted in Fig.~\ref{fig:unet} for each possible output of the synchronization-to-interference \gls{cnn} block. In other words, we implement a ``conditional separation" block for each possible time-shift of the interference. 
If explicit synchronization-to-interference is not used, the raw unprocessed mixture is (always) fed into to the same \gls{dnn} separation block.

 The training set is processed as such to yield a labeled dataset (mixture $\rvecy$ and ground-truth reference signal $\rvecs$). As a loss function, we use the empirical \gls{mse}. For full implementation details, see our Github repository.\footnote{\label{fn:github}\url{https://github.com/RFChallenge/SCSS_Sync}}
 %
 \section{Numerical Results}\label{sec:num}
We generate synthetic mixtures $\rvecy$ where the \gls{soi} bears \gls{qpsk} symbols using a root-raised cosine pulse-shaping filter with roll-off factor $0.5$, spanning $8$ \gls{qpsk} symbols, and with an oversampling factor $16$. The interference is an \gls{ofdm} signal. We generate an \gls{ofdm} signal with symbols of length $80$, bearing $16$--\gls{qam} symbols, with a \gls{fft} size of $64$, and a cyclic-prefix of length $16$. Details on the signals generation process are provided in the Github repository.\footnoteref{fn:github}

\subsection{The Potential Gain of Explicit Synchronization with DNNs}\label{sec:expl_sync_num}
We now compare the performance of the \gls{dnn} approach illustrated in Fig.~\ref{fig:system} with the performance achieved by classical methods for detection and interference rejection, i.e., \gls{mf} and the \gls{lmmse} estimator $\widehat{\rvecs}_{\lmmse}$ given in~\eqref{lmmsedef}, and by our proposed ``synchronized" \gls{qlmmse} estimator $\widehat{\rvecs}_{\cnnqlmmse}$ given in~\eqref{eq:cnn-qlmmse}. For the \gls{cnn}-based synchronization-to-interference methods (Section~\ref{sec:synchviacnn}), we input $640$ samples of the mixture $\rvecy$ to the \gls{cnn}. The input size to the separation U-Net is $N=10240$. 
 
\begin{figure}[t!]     
    \centering
    \includegraphics[width=\columnwidth]{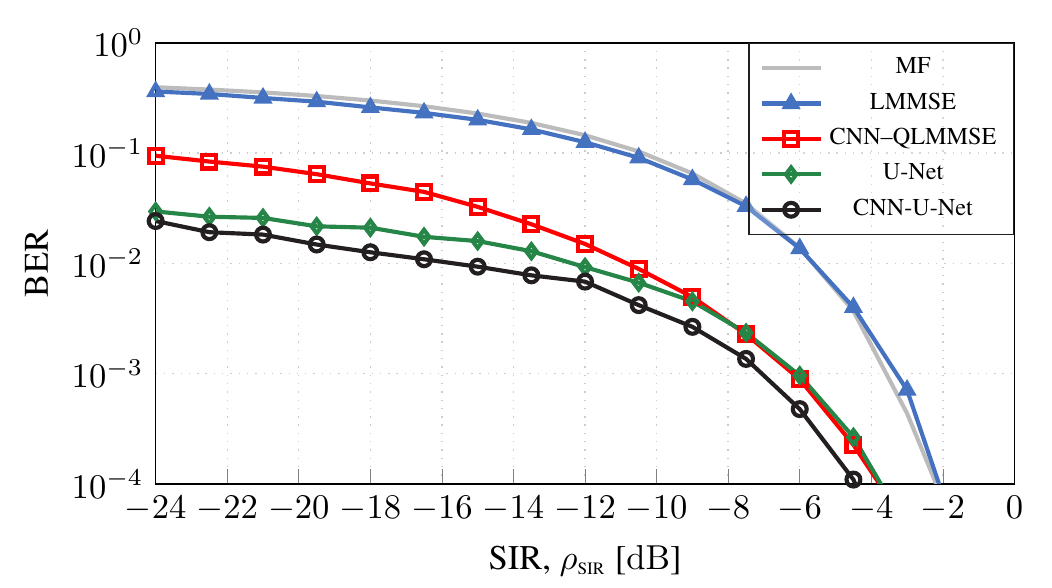}
    \caption{\gls{ber} as a function of the SIR for \gls{mf} detection; \gls{lmmse} and \gls{qlmmse} interference rejection (blocks of length 320); and the data-driven U-Net approach with and without synchronization to the interference.}
    \label{fig:Unet_sync} 
 \end{figure}

In Fig.~\ref{fig:Unet_sync}, we compare the performance in terms of \gls{ber} as a function of the \gls{sir} in a noiseless setting. Specifically, we depict in gray the \gls{mf} approach. 
 In blue, we depict the \gls{lmmse} ($\widehat{\rvecs}_{\lmmse}$ in~\eqref{lmmsedef}) computed using blocks of length $320$.\footnote{For non-stationary processes, the required inversion of $\matC_{yy}$ is computationally impractical for large $N$, as it is generally of complexity $\mathcal{O}(N^{3})$.} In red, we depict the \gls{cnn}--\gls{qlmmse} approach ($\widehat{\rvecs}_{\cnnqlmmse}$ in~\eqref{eq:cnn-qlmmse}), also using blocks of length $320$. Here, we explicitly synchronize to the interference signal, and exploit this to obtain ``aligned statistics" \eqref{alignedstatistics} for each possible time-shift $\rndk_b$. In green, we depict the performance of the U-Net approach when there is no explicit synchronization, i.e., the ``Synchronization CNN" block in Fig.~\ref{fig:system} is removed. Finally, we depict in black the \gls{dnn} approach including both the synchronization and separation blocks, as described in Fig.~\ref{fig:system}, denoted as \gls{cnn}--U-Net. Every described approach includes a last \gls{mf} step before hard decoding based on the minimum Euclidean distance rule. 
 
 As can be observed, by only applying a \gls{mf} to the received signal $\rvecy$, which is optimal under white Gaussian noise, we do not exploit any temporal structure of the (non-Gaussian) interference. Hence, as expected, we obtain the worst performance. It is also evident that the \gls{lmmse} approach---optimal for Gaussian signals---without explicit alignment of the signal statistics via synchronization, is unable to exploit the underlying temporal nonstationarities, and accordingly yields approximately the performance obtained by only applying a \gls{mf} to the received signal $\rvecy$. However, by explicitly synchronizing to the interference signal using the \gls{cnn} described in Section~\ref{sec:synchviacnn}, we can now use the \emph{conditional} covariance of the interference for each possible time-shift $\rndk_b$ to obtain $\widehat{\rvecs}_{\cnnqlmmse}$, which already leads to a significant performance gain. For example, for a \gls{ber} of $10^{-3}$, the \gls{cnn}--\gls{qlmmse} approach requires an \gls{sir} of $-6$\dB, while the \gls{mf} and the \gls{lmmse} approaches require $-4$\dB. Even though by explicitly synchronizing to the interference we can obtain significant gains, we recall that by using (quasi-)linear processing we can only exploit up to (conditional) second order statistics. 
 
 Since we consider digital communication signals, further gains can be achieved by exploiting high-order statistics and the ``discrete nature" of these signals. This is precisely achieved by our proposed \gls{dnn}-based approaches (green and black). First, it is observed that a U-Net without prior explicit synchronization already outperforms the \gls{cnn}--\gls{qlmmse} approach for most of the considered \gls{sir} values. 
 The performance of the U-Net is further improved with explicit synchronization, using the block described in Fig.~\ref{fig:cnn_sync}, as shown in Fig.~\ref{fig:system}. In this case, a \gls{ber} of $10^{-2}$ is obtained at an \gls{sir} level of $-17$\dB, while the U-Net without explicit synchronization requires $-12$\dB, and the \gls{cnn}--\gls{qlmmse} approach requires $-10.5$\dB. Thus, for a given architecture with limited capacity (parametrization power), decoupling synchronization and separation can lead to considerable gains, which enables reliable communication in the presence of strong interference.
\subsection{Gains from Explicit-Synchronization-Free Architecture}\label{sec:impl_high_dim}
As mentioned in Section~\ref{sec:int_rej_DNNs}, a plausible interpretation of Lemma~\ref{lemma1} is the following. When the input mixtures are sufficiently long, an explicit-synchronization-based architecture may not be required (or even provide superior performance), since the data is “very informative” with respect to the underlying time-shift. This essentially makes direct separation (i.e., an “implicit” synchronization approach) potentially preferable. Our best result up to date is achieved by directly inputting mixtures of length $N=40960$ to the U-Net depicted in Fig.~\ref{fig:unet}.

\begin{figure}[t!]     
    \centering
    \includegraphics[width=\columnwidth]{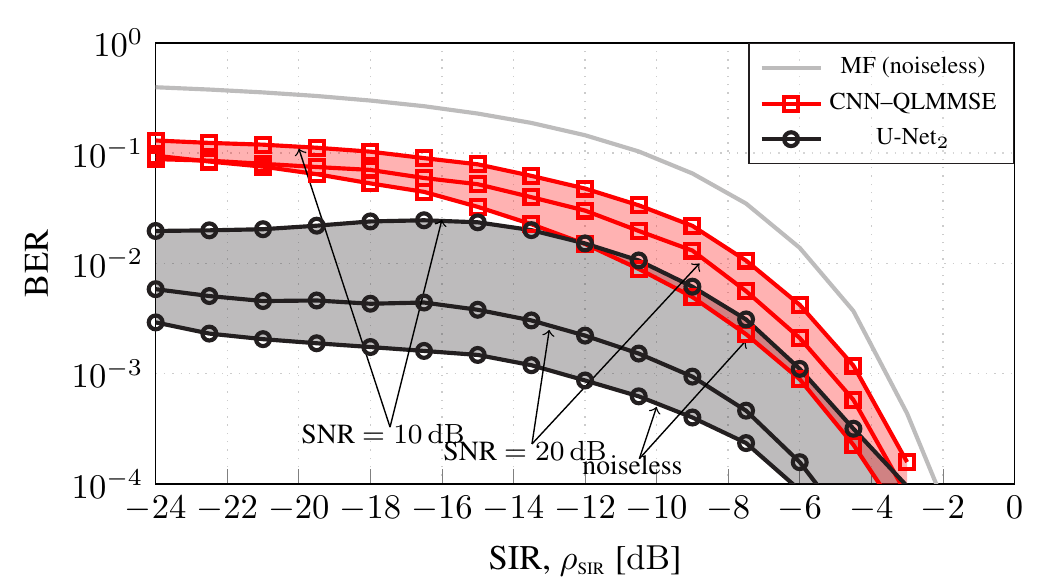}
    \caption{\gls{ber} as a function of the SIR for the noiseless case and SNR$=\{10,20\}$\dB, for \gls{mf} (noiseless only), \gls{qlmmse} interference rejection (blocks of length $320$), and the CNN--U-Net approach described in Fig.~\ref{fig:system}.}
    \label{fig:Unet_sync_noisy} 
\end{figure}

Fig.~\ref{fig:Unet_sync_noisy} shows the performance of the U-Net scheme described in Fig.~\ref{fig:unet} (U-Net$_2$) where we input two replicas of the mixture $\rvecy$, which provides the first layer with more diversity. We consider three different \gls{snr} levels of white Gaussian noise: $\rho_{\snr}\in\{10,20,\infty\}$\dB.
Specifically, we compare the performance of the \gls{dnn} solution with the performance of the \gls{cnn}--\gls{qlmmse} approach (computed using blocks of length $320$) and the \gls{mf} approach, which is only plotted for the noiseless case for the sake of clarity.
Clearly, for all \gls{snr} levels, the U-Net$_2$ approach outperforms the \gls{cnn}--\gls{qlmmse} and \gls{mf} approaches. However, as expected, the smaller the \gls{snr}, the smaller the gap between them. For example, for a \gls{ber} of $10^{-3}$, the gain of the U-Net$_2$ approach compared to the \gls{cnn}--\gls{qlmmse} is roughly $7$\dB in the noiseless case, $4$\dB for an \gls{snr}$=20$\dB, and $1.8$\dB for an \gls{snr}$=10$\dB.
 %
\section{Conclusions and Outlook}\label{sec:conclusions}
 We study the \gls{scss} problem with a focus on its application to interference rejection in digital communication. For Gaussian signals, we prove that a decoupled system architecture of synchronization followed by separation is asymptotically optimal in the \gls{mmse} sense. Consequently, since the optimal system can be impractical for implementation purposes, we propose a computationally attractive alternative with negligible performance loss relative to the optimal system. For (non-Gaussian) signals, we demonstrate in simulations that the proposed \gls{dnn}-based data-driven approach can exploit the underlying temporal structures of the signals, thus leading to significant gains in terms of \gls{ber}, and in particular, outperforms classical methods.
 Extensions of this work should focus on understanding how and when to use explicit synchronization  in the context of \gls{scss} with \glspl{dnn}.  

 
 
 


 

\appendices

\section{Proof of Lemma \ref{lemma1}}\label{app:proof_lemma1}

To prove Lemma \ref{lemma1}, we shall use the following lemma.
\begin{lemma}\label{lemma2}
For $\psi_N(\rvecy,k)$, in Definition \ref{definition1} (\gls{usc}), we have,
\begin{align}\label{conditionlemma2}
    &\Exop\left[e^{\tau\psi_N(\rvecy,\rndk_b)}\right]=\left(1-\frac{\tau}{N}\right)^{-N}\cdot e^{-\tau}, \;\; \forall \tau<N.
\end{align}
\end{lemma}
\begin{IEEEproof}[Proof of Lemma \ref{lemma2}]
First, recall $\rvecy|\rndk_b\sim\mathcal{CN}(\veczero,\matC_{yy}(\rndk_b))$, where $\matC_{yy}(\rndk_b)=\matC_{ss}(0)+\matC_{vv}(\rndk_b)$. Using the Cholesky decomposition, we write $\matC_{yy}(\rndk_b)\triangleq\matGamma_y(\rndk_b)\herm{\matGamma}_y(\rndk_b)$, where $\matGamma_y(\rndk_b)\in\complexset^{N\times N}$. Then, conditioned on $\rndk_b$, we have
\begin{IEEEeqnarray}{rCl}
    \psi_N(\rvecy,\rndk_b)+1 & = & \frac{1}{N}\herm{\rvecy}\matC^{-1}_{yy}(\rndk_b)\rvecy
    \\
    & = & \frac{1}{N}\herm{\rvecy}\herminv{\matGamma}_y(\rndk_b)\inv{\matGamma}_y(\rndk_b)\rvecy
    \\
    & = & \frac{1}{N}\herm{\Big(\underbrace{\inv{\matGamma}_y(\rndk_b)\rvecy}_{\triangleq\rvecu(\rndk_b)}\Big)}\underbrace{\inv{\matGamma}_y(\rndk_b)\rvecy}_{=\rvecu(\rndk_b)}
    \\
    & = &\frac{1}{N}\|\rvecu(\rndk_b)\|_2^2,\label{normalizedwhitevector}
\end{IEEEeqnarray}
where $\rvecu(\rndk_b)|\rndk_b\sim\mathcal{CN}(\veczero,\matI)$ is a white Gaussian vector. Thus,
\begin{IEEEeqnarray}{rCl}
  \hspace{-0.3cm}\Exop\left[e^{\tau\psi_N(\rvecy,\rndk_b)}\right] & = & \Exop\left[\Exop\left[e^{\tau\psi_N(\rvecy,\rndk_b)}|\rndk_b\right]\right]\label{lemma2trans1}\\
  &=&\Exop\left[\Exop\left[e^{\tau\left(\frac{1}{N}\|\rvecu(\rndk_b)\|_2^2-1\right)}|\rndk_b\right]\right]\\
  &=&\Exop\left[\Exop\left[e^{\frac{\tau}{N}\sum_{n=1}^N|\rndu_n(\rndk_b)|^2}|\rndk_b\right]\right]e^{-\tau}\\
  &=&\Exop\left[\prod_{n=1}^N\Exop\left[e^{\frac{\tau}{2N}|\sqrt{2}\rndu_n(\rndk_b)|^2}|\rndk_b\right]\right]e^{-\tau}\label{lemma2trans2}\\
  &\overset{\forall \tau<N}{=}&\Exop\left[\prod_{n=1}^N\left(1-\frac{\tau}{N}\right)^{-1}\right]e^{-\tau}\label{lemma2trans3}\\
  &=&\left(1-\frac{\tau}{N}\right)^{-N}\cdot e^{-\tau},
\end{IEEEeqnarray}
where we have used the law of total expectation in \eqref{lemma2trans1}; the conditional statistical independence of the elements of $\rvecu(\rndk_b)$ (given $\rndk_b$) in \eqref{lemma2trans2}; the fact that $\{|\sqrt{2}\rndu_n(\rndk_b)|^2\sim\chi^2_2\}_{n=1}^N$, namely all the squared absolute-valued elements of $\rvecu(\rndk_b)$, given $\rndk_b$, are chi-squared random variables with two degrees of freedom; and, accordingly, that the moment generating function of a random variable $\rndq\sim\chi^2_2$ is $\Exop[e^{\widetilde{\tau}\rndq}]=(1-2\widetilde{\tau})^{-1}$, for all $\widetilde{\tau}<\frac{1}{2}$, in \eqref{lemma2trans3}, where in our case $\widetilde{\tau}=\tau/2N$, hence the condition on $\tau$ in \eqref{lemma2trans2}
\end{IEEEproof}
Equipped with Lemma \ref{lemma2}, we now prove Lemma \ref{lemma1}.

By definition, the \gls{map} estimator has the lowest error probability. Therefore, to show \eqref{polydecayoferrprob}, it is sufficient to show that there exists another estimator of $\rndk_b$, whose error probability is $o(N^{-\alpha})$ for any finite $\alpha\in\reals_+$, independent of $N$. For this, let us consider the estimator,
\begin{equation}\label{simpleestimatorfork_b}
    \widehat{\rndk}_{b}\triangleq \argmin_{m\in\setS_{K_b}} \left|\psi_N(\rvecy,m)\right|.
\end{equation}
In words, as $N\to\infty$, the error probability of \eqref{simpleestimatorfork_b} is governed by how far is $|\psi_N(\rvecy,\rndk_b)|$ from zero, since from the \gls{usc}, $\nexists k\in\setS_{K_b}\backslash\rndk_b: \lim_{N\to\infty} |\psi_N(\rvecy,k)|=0$, whereas
\begin{IEEEeqnarray}{rCl}
  \lim_{N\to\infty}\psi_N(\rvecy,\rndk_b) & = & \Exop\left[\psi_N(\rvecy,\rndk_b)\right]\label{limconvergetomean}
  \\
  & = & \Exop\left[\Exop\left[\psi_N(\rvecy,\rndk_b)|\rndk_b\right]\right]
  \\
  & = & \frac{1}{N}\Exop\left[\Exop\left[\|\rvecu(\rndk_b)\|_2^2|\rndk_b\right]\right]-1=0,\label{lastlinelim2mean}
\end{IEEEeqnarray}
where we have used \eqref{normalizedwhitevector}, $\rvecu(\rndk_b)|\rndk_b\sim\mathcal{CN}(\veczero,\matI)$, and \eqref{limconvergetomean} follows from the fact that $\Varop(\psi_N(\rvecy,\rndk_b))=1/N$, which can be shown in a similar fashion to \eqref{limconvergetomean}--\eqref{lastlinelim2mean}.

Formally, the error probability of this estimator is given by,
\begin{equation}
    \prob{\widehat{\rndk}_{b}\neq\rndk_{b}}=\prob{|\psi_N(\rvecy,\rndk_b)|>\min_{m\in\setS_{K_b}\backslash\rndk_b}|\psi_N(\rvecy,m)|}.
\end{equation}
We now show that the probability that $\psi_N(\rvecy,\rndk_b)$ is bounded away from zero decreases in the desired rate. Clearly, for any $a>0$, we have
\begin{IEEEeqnarray}{rCl}\label{split2twoprobs}
    \prob{|\psi_N(\rvecy,\rndk_b)|>a}& = &\prob{\psi_N(\rvecy,\rndk_b)>a}\\
    && + \: \prob{\psi_N(\rvecy,\rndk_b)<-a}.
\end{IEEEeqnarray}

Using the Chernoff bound, we have
\begin{align}
    &\prob{\psi_N(\rvecy,\rndk_b)>a} \leq \Exop\left[e^{t\psi_N(\rvecy,\rndk_b)}\right]e^{-ta}\triangleq B_1(t,a),\label{cherupperbound1}\\
    &\prob{\psi_N(\rvecy,\rndk_b)<-a} \leq  \Exop\left[e^{-t\psi_N(\rvecy,\rndk_b)}\right]e^{-ta}\triangleq B_2(t,a).\label{cherupperbound2}
\end{align}
Using Lemma \ref{lemma2}, it follows that
\begin{align}
    &B_1(t,a)=\left(1-\frac{t}{N}\right)^{-N}\cdot e^{-t(1+a)},\quad \forall t<N,\\
    &B_2(t,a)=\left(1+\frac{t}{N}\right)^{-N}\cdot e^{t(1-a)}, \quad \forall t>-N.
\end{align}
Minimizing $B_1(t,a)$ and $B_2(t,a)$ with respect to $t$ and choosing $a=N^{-(0.5-\epsilon)}$ for some $0<\epsilon<0.5$, we obtain
\begin{align}
    \min_{t<N}B_1(t,N^{-(0.5-\epsilon)})&= \left(1+\frac{1}{N^{(0.5-\epsilon)}}\right)^{N}e^{-N^{0.5+\epsilon}}
    \\
    &\triangleq B^*_1[N],\\
    \min_{t>-N}B_2(t,N^{-(0.5-\epsilon)})&= \left(1-\frac{1}{N^{(0.5-\epsilon)}}\right)^{N}e^{N^{0.5+\epsilon}}
    \\
    &\triangleq B^*_2[N].
\end{align}
Finally, as for any $\alpha\in\reals_+$ and any $\delta>0$ independent of $N$,
\begin{align}
    \lim_{N\to\infty} N^{\alpha+\delta}B^*_1[N] = \lim_{N\to\infty} N^{\alpha+\delta}B^*_2[N] = 0,
\end{align}
it follows that for any $\alpha\in\reals_+$ independent of $N$,
\begin{align}
    \prob{|\psi_N(\rvecy,\rndk_b)|>\frac{1}{N^{0.5-\epsilon}}}&= o\left(\frac{1}{N^{\alpha}}\right)
    \\
    \Longrightarrow \; \prob{\widehat{\rndk}_{b}\neq\rndk_{b}}&=o\left(\frac{1}{N^{\alpha}}\right)\label{bigOforanyalpha}.
\end{align}
\section{Proof of Theorem \ref{theorem1}}\label{app:proof_th}
From Lemma \ref{lemma1}, we have the following corollary.
\begin{corollary}\label{corollary1}
Using \eqref{polydecayoferrprob}, we have
\begin{align}
    &\Exop\left[\prob{\widehat{\rndk}^{\map}_{b}\neq\rndk_{b}|\rvecy}\right]=o\left(\frac{1}{N^{\alpha}}\right),\label{expdecayoferrprobexpectationV0}\\
    &\Exop\left[\prob{\widehat{\rndk}^{\map}_{b}=k|\rvecy}\right]=o\left(\frac{1}{N^{\alpha}}\right), \; \forall k\in\setS_{K_b}\backslash \rndk_b.\label{expdecayoferrprobexpectation}
\end{align}
\end{corollary}

The roadmap for the proof of the theorem is as follows:
\begin{itemize}
    \item \underline{Step 1}: Express the optimality gap between the \gls{mmse} and \gls{map}-based \gls{qlmmse} estimators as a function of the error probability of the \gls{map} synchronizer $\widehat{\rndk}^{\map}_{b}$.
    \item \underline{Step 2}: Express the \gls{mmse} \eqref{mmsedef} as a sum of the \gls{map}-based \gls{qlmmse} \eqref{qlmmsedef} and the expected squared norm of the optimality gap, also known as the ``regret".
    \item \underline{Step 3}: Show that the regret is upper bounded by terms that decay polynomially fast, for any fixed polynomial rate (using Lemma \ref{lemma1}).
\end{itemize}
We now prove Thoerem \ref{theorem1}.
Let us write the the \gls{mmse} estimator \eqref{mmsedefest}, explicitly, using \eqref{ourmmse}, in terms of the \gls{map}-based \gls{qlmmse} estimator \eqref{mapqlmmse}, as (recall $\rndk_s=0$, by assumption),
\begin{equation}\label{mmseintermsofmapqlmmse}
\begin{aligned}
    \widehat{\rvecs}_{\mmse}=&\sum_{m_b=1}^{K_b}\prob{\rndk_b=m_b | \rvecy}\widehat{\rvecs}_{\lmmse}(m_b)\\
    =&\underbrace{\sum_{\substack{m_b=1 \\ m_b\neq \widehat{\rndk}^{\map}_{b}}}^{K_b}\prob{\rndk_b=m_b | \rvecy}\widehat{\rvecs}_{\lmmse}(m_b)}_{\triangleq\rvecdelta(\rvecy)}\\
    &\quad\quad\;+\prob{\rndk_b=\widehat{\rndk}^{\map}_{b} | \rvecy}\widehat{\rvecs}_{\lmmse}(\widehat{\rndk}^{\map}_{b}).\\
\end{aligned}
\end{equation}
Using \eqref{mmseintermsofmapqlmmse}, we define the optimality gap (vector),
\begin{align}
    \rvecDelta(\rvecy)&\triangleq \widehat{\rvecs}_{\mmse} - \widehat{\rvecs}_{\lmmse}(\widehat{\rndk}^{\map}_{b})=\widehat{\rvecs}_{\mmse} -\widehat{\rvecs}_{\qlmmse}\label{transition1}\\
    &=\rvecdelta(\rvecy)-\prob{\rndk_b\neq\widehat{\rndk}^{\map}_{b} | \rvecy}\widehat{\rvecs}_{\qlmmse}.
\end{align}


Let us proceed to the second step of the proof. For shorthand, let $\rvece_{\qlmmse}\triangleq\widehat{\rvecs}_{\qlmmse}-\rvecs$, and let us first write the \gls{mmse} in terms of the estimation error $\rvece_{\qlmmse}$ and the optimality gap $\rvecDelta(\rvecy)$ as,
\begin{align}
    \Exop\left[\|\widehat{\rvecs}_{\mmse}-\rvecs\|_2^2\right]&=\Exop\left[\|\widehat{\rvecs}_{\mmse}-\widehat{\rvecs}_{\qlmmse}+\widehat{\rvecs}_{\qlmmse}-\rvecs\|_2^2\right]\label{transition1equality}\\
    &=\varepsilon^2_{\qlmmse}(N)-\Exop\left[\|\rvecDelta(\rvecy)\|_2^2\right],\label{transition1equality2}
\end{align}
where we have used \eqref{transition1} in \eqref{transition1equality}, and the well-known orthogonality property of the estimation error in MMSE estimation to any function of the measurements in \eqref{transition1equality2}. Expanding the first term, we have,
\begin{equation}\label{bigdeltaexplicitly}
\begin{aligned}
    \Exop\left[\|\rvecDelta(\rvecy)\|_2^2\right]&=\\
    \Exop\left[\|\rvecdelta(\rvecy)\|_2^2\right]&+\Exop\left[\prob{\rndk_b\neq\widehat{\rndk}^{\map}_{b} | \rvecy}^2\|\widehat{\rvecs}_{\qlmmse}\|_2^2\right]\\
    &-2\Re\left\{\Exop\left[\prob{\rndk_b\neq\widehat{\rndk}^{\map}_{b} | \rvecy}\herm{\rvecdelta}(\rvecy)\widehat{\rvecs}_{\qlmmse}\right]\right\}.
\end{aligned}
\end{equation}
We now show that (the magnitude of) each of the terms in \eqref{bigdeltaexplicitly} is bounded. It will then follow that the expected squared norm of the optimality gap, $\Exop\left[\|\rvecDelta(\rvecy)\|_2^2\right]$, is also bounded.

Starting with the first term in \eqref{bigdeltaexplicitly}, we have,
\begin{align}
    &\Exop\left[\|\rvecdelta(\rvecy)\|_2^2\right]=\sum_{n=1}^N\Exop\left[\delta_n^2(\rvecy)\right]=\\
    &\sum_{n=1}^N\Exop\left[\left(\sum_{\substack{m_b=1 \\ m_b\neq \widehat{\rndk}^{\map}_{b}}}^{K_b}\prob{\rndk_b=m_b | \rvecy}\widehat{\rnds}_{\lmmse,n}(m_b)\right)^2\right].\label{sumofexpectations}
\end{align}
Focusing on one element of the sum in \eqref{sumofexpectations}, we have,
\begin{align}
    &\Exop\left[\left(\sum_{\substack{m_b=1 \\ m_b\neq \widehat{\rndk}^{\map}_{b}}}^{K_b}\prob{\rndk_b=m_b | \rvecy}\widehat{\rnds}_{\lmmse,n}(m_b)\right)^2\right]\leq\label{CSineq1}\\
    &\sum_{\substack{m_1=1 \\ m_1\neq \widehat{\rndk}^{\map}_{b}}}^{K_b}\sum_{\substack{m_2=1 \\ m_2\neq \widehat{\rndk}^{\map}_{b}}}^{K_b}\Exop\left[\prob{\rndk_b=m_1 | \rvecy}^2\prob{\rndk_b=m_2 | \rvecy}^2\right]^{\frac{1}{2}}\cdot\nonumber\\
    &\quad\quad\quad\quad\;\;\,\quad\quad\Exop\left[\widehat{\rnds}^2_{\lmmse,n}(m_1)\widehat{\rnds}^2_{\lmmse,n}(m_2)\right]^{\frac{1}{2}}\leq\label{CSineq2}\\
    &\sum_{\substack{m_1=1 \\ m_1\neq \widehat{\rndk}^{\map}_{b}}}^{K_b}\sum_{\substack{m_2=1 \\ m_2\neq \widehat{\rndk}^{\map}_{b}}}^{K_b}{\Exop\left[\prob{\rndk_b=m_1 | \rvecy}^4\right]}^{\frac{1}{4}}{\Exop\left[\prob{\rndk_b=m_2 | \rvecy}^4\right]}^{\frac{1}{4}}\cdot\nonumber\\
    &\quad\quad\quad\quad\;\;\,\quad\quad\Exop\left[\widehat{\rnds}^2_{\lmmse,n}(m_1)\widehat{\rnds}^2_{\lmmse,n}(m_2)\right]^{\frac{1}{2}}\leq\label{trivialprobinequality}\\
    &\sum_{\substack{m_1=1 \\ m_1\neq \widehat{\rndk}^{\map}_{b}}}^{K_b}\sum_{\substack{m_2=1 \\ m_2\neq \widehat{\rndk}^{\map}_{b}}}^{K_b}{\underbrace{\Exop\left[\prob{\rndk_b=m_1 | \rvecy}\right]}_{o\left(\frac{1}{N^{4\alpha}}\right)}}^{\frac{1}{4}}{\underbrace{\Exop\left[\prob{\rndk_b=m_2 | \rvecy}\right]}_{o\left(\frac{1}{N^{4\alpha}}\right)}}^{\frac{1}{4}}\cdot\label{usingcorollary1}\\
    &\quad\quad\quad\quad\;\;\,\quad\quad\underbrace{\Exop\left[\widehat{\rnds}^2_{\lmmse,n}(m_1)\widehat{\rnds}^2_{\lmmse,n}(m_2)\right]^{\frac{1}{2}}}_{\mathcal{O}\left(1\right)}=\\
    &\quad\quad\quad\quad\;\;\,\quad\quad\;o\left(\frac{1}{N^{\alpha}}\right),
\end{align}
where we have used the Cauchy-Schwarz inequality repeatedly in \eqref{CSineq1} and \eqref{CSineq2},
the following (almost trivial) observation,
\begin{equation}\label{trivialboundonprobability}
    \prob{\rvecz=\vecz}^{\beta}\leq \prob{\rvecz=\vecz}, \quad \forall\beta\geq1,
\end{equation}
in \eqref{trivialprobinequality}, and \eqref{expdecayoferrprobexpectation} in \eqref{usingcorollary1}.
Since \eqref{sumofexpectations} is a sum of $N$ terms as in \eqref{CSineq1}, we obtain
\begin{equation}\label{firstterminequestablished}
    \Exop\left[\|\rvecdelta(\rvecy)\|_2^2\right]=o\left(\frac{1}{N^{\alpha-1}}\right).
\end{equation}
Moving to the second term in \eqref{bigdeltaexplicitly}, we have,
\begin{align}
    &\Exop\left[\prob{\rndk_b\neq\widehat{\rndk}^{\map}_{b} | \rvecy}^2\|\widehat{\rvecs}_{\qlmmse}\|_2^2\right]\leq\label{trans12ndterm}\\
    &\Exop\left[\prob{\rndk_b\neq\widehat{\rndk}^{\map}_{b} | \rvecy}\|\widehat{\rvecs}_{\qlmmse}\|_2^2\right]\leq\label{trans22ndterm}\\
    &\Exop\left[\prob{\rndk_b\neq\widehat{\rndk}^{\map}_{b} | \rvecy}^2\right]^{\frac{1}{2}}\Exop\left[\|\widehat{\rvecs}_{\qlmmse}\|_2^4\right]^{\frac{1}{2}}\leq\label{trans32ndterm}\\
    &\,{\underbrace{\Exop\left[\prob{\rndk_b\neq\widehat{\rndk}^{\map}_{b} | \rvecy}\right]}_{o\left(\frac{1}{N^{2\alpha}}\right)}}^{\frac{1}{2}}{\underbrace{\Exop\left[\|\widehat{\rvecs}_{\qlmmse}\|_2^4\right]}_{\mathcal{O}\left(N\right)}}^{\frac{1}{2}}=o\left(\frac{1}{N^{\alpha-1}}\right)\label{trans42ndterm},
\end{align}
where we have used \eqref{trivialboundonprobability} in \eqref{trans12ndterm} and \eqref{trans32ndterm}, the Cauchy-Schwarz inequality in \eqref{trans22ndterm}, and \eqref{expdecayoferrprobexpectationV0} in \eqref{trans42ndterm}. As for the magnitude of the last term in \eqref{bigdeltaexplicitly}, we similarly obtain,
\begin{align}
    &\left|\Re\left\{\Exop\left[\prob{\rndk_b\neq\widehat{\rndk}^{\map}_{b} | \rvecy}\herm{\rvecdelta}(\rvecy)\widehat{\rvecs}_{\qlmmse}\right]\right\}\right|\leq\\
    &\left|\Exop\left[\prob{\rndk_b\neq\widehat{\rndk}^{\map}_{b} | \rvecy}\herm{\rvecdelta}(\rvecy)\widehat{\rvecs}_{\qlmmse}\right]\right|\leq\label{trans13rdterm}\\
    &\;\Exop\left[\prob{\rndk_b\neq\widehat{\rndk}^{\map}_{b} | \rvecy}^2\right]^{\frac{1}{2}}\Exop\left[\left|\herm{\rvecdelta}(\rvecy)\widehat{\rvecs}_{\qlmmse}\right|^2\right]^{\frac{1}{2}}\leq\label{trans23rdterm}\\
    &\;{\underbrace{\Exop\left[\prob{\rndk_b\neq\widehat{\rndk}^{\map}_{b} | \rvecy}\right]}_{o\left(\frac{1}{N^{2\alpha}}\right)}}^{\frac{1}{2}}{\underbrace{\Exop\left[\left|\herm{\rvecdelta}(\rvecy)\widehat{\rvecs}_{\qlmmse}\right|^2\right]}_{\mathcal{O}\left(N\right)}}^{\frac{1}{2}}=\label{tran33rdterm}\\
    &\;o\left(\frac{1}{N^{\alpha-1}}\right),\label{bound3rdterm}
\end{align}
where we have used, again, the Cauchy-Schwarz inequality in \eqref{trans13rdterm}, \eqref{trivialboundonprobability} in \eqref{trans23rdterm}, and \eqref{expdecayoferrprobexpectationV0} in \eqref{tran33rdterm}. We note in passing that the term on the right in \eqref{tran33rdterm} may be bound
more tightly, but this is not necessary for the following steps of this proof.

We have established upper bounds on the magnitudes of the terms in \eqref{bigdeltaexplicitly}. Hence, using \eqref{firstterminequestablished}, \eqref{trans42ndterm} and \eqref{bound3rdterm}, we now have
\begin{align}\label{boundonbigdelta}
    &\Exop\left[\|\rvecDelta(\rvecy)\|_2^2\right]=o\left(\frac{1}{N^{\alpha-1}}\right),
\end{align}
which, together with \eqref{transition1equality2}, yields
\begin{equation}\label{equalityuptobigON}
    \varepsilon^2_{\mmse}(N)=\varepsilon^2_{\qlmmse}(N)+o\left(\frac{1}{N^{\alpha-1}}\right).
\end{equation}
By the definition of the \gls{mmse} estimator, the (trivial) upper bound
\begin{equation}\label{upperboundsandwitch}
    \varepsilon^2_{\mmse}(N)\leq\varepsilon^2_{\qlmmse}(N) \; \Longrightarrow \; \frac{\varepsilon^2_{\mmse}(N)}{\varepsilon^2_{\qlmmse}(N)}\leq 1
\end{equation}
holds for any $N\in\posnaturals$. Therefore, and since \eqref{equalityuptobigON} hold for any $\alpha\in\reals_+$, we can always choose some $\alpha$ to have
\begin{equation}
    \frac{\varepsilon^2_{\mmse}(N)}{\varepsilon^2_{\qlmmse}(N)}=1-o\left(\frac{1}{N^{\alpha}}\right),
\end{equation}
where we used $\varepsilon^2_{\qlmmse}(N)=\mathcal{O}(N)$, proving the theorem.

\bibliographystyle{IEEEtran}

\end{document}